\documentclass[twocolumn,showpacs,preprintnumbers,amsmath,amssymb,aps]{revtex4-1}
\usepackage{graphicx}
\usepackage{dcolumn}
\usepackage{bm}
\usepackage{float}
\usepackage{color}
\usepackage{soul}
\usepackage{mathtools}
\usepackage{ragged2e}
\begin{document}

\title{A single measurement scheme for quantum work statistics based on coherent or squeezing state}

\author{Bao-Ming Xu$^{1}$}
\email{xbmv@bit.edu.cn}
\author{Jian Zou$^{2}$}
\author{Zhan-Chun Tu$^{3}$}
\affiliation{$^{1}$Shandong Key Laboratory of Biophysics, Institute of Biophysics, Dezhou University, Dezhou 253023, China}%
\affiliation{$^{2}$School of Physics, Beijing Institute of Technology, Beijing 100081, China}
\affiliation{$^{3}$Department of Physics, Beijing Normal University, Beijing 100875, China}

\date{Submitted \today}

\begin{abstract}
In order to investigate the role of initial quantum coherence in work probability distribution, it is necessary to explicitly consider a concrete measurement apparatus to record work rather than implicitly appealing to perform an energy measurement. In this paper, we consider a harmonic oscillator with coherent or squeezing state as measurement apparatus, and propose a unified framework of quantum work statistics for arbitrary initial state. We find that work is proportional to the change of the real part of coherent state parameter, i.e., quantum work can be estimated by the coherent state parameter. The resulting work probability distribution includes the initial quantum coherence, and can be reduced to result of traditional two projective energy measurement scheme (TPM) by squeezing the state of the harmonic oscillator. As an application, we consider a driven two-level system and investigate the effects of driving velocity on work statistics. We find that only when the driving velocity matches the transition frequency of the system can initial quantum coherence play an important role.
\end{abstract}

\maketitle

\section{Introduction}
Triggered by recent advances in the coherent manipulation of elementary quantum systems \cite{Bloch2008,Kippenberg2007,Aspelmeyer2014}, out-of-equilibrium quantum thermodynamics has been arouse enormous research interest recently \cite{Adesso2018}. The concept of work is one of the cornerstones of thermodynamics. To extend it and its fluctuation theorems to the quantum domain is of fundamental importance in quantum thermodynamics. Different from the classical case, it is very hard to define work in the quantum regime, because work is not an observable \cite{Talkner2007a}. Traditionally, to determine the work in the quantum regime one needs to perform two projective energy measurement (TPM) at the beginning and the end of the external protocol \cite{Esposito2009,Campisi2011}. Based on TPM, the quantum extension of fluctuation theorems is obtained, see the reviews \cite{Esposito2009,Campisi2011} for detail discussion, and these fluctuation theorems have been experimentally verified in various systems \cite{Batalhao2014,Huber2008,An2015,Hoang2018,Xiong2018,Cerisola2017,Naghiloo2018}. However, the initial quantum coherence is destroyed by the first measurement, and therefore the work fluctuation relation obtained by TPM is not ``quantum" to some extent. And due to the severe impact of projective measurement on the system dynamics, the first law of thermodynamics can not be satisfied. 

In order to include the effects of initial quantum coherence, a lot of efforts have been made. Replacing projective measurement by weak measurement \cite{Watanabe2014,Talkner2016}, a part of quantum coherence can be preserved, but the impact of measurement can not be eliminated. Based on the fact that work is a process accumulated quantity, Refs. \cite{Solinas2013,Venkatesh2015,Miller2017} defined the work by the integral of the injected power during the evolution, and obtained the quantum work distribution including initial quantum coherence. Analogy to classical trajectory in phase space, Sampaio \textit{et al}. gave the quantum trajectory using quantum Hamilton-Jacobi theory, and defined work as the power integrated along the trajectory \cite{Sampaio2018}. The resulting work distribution is valid for any quantum evolution, including those with coherence in the energy basis. Notably, all the work distribution above can not satisfy the first law of thermodynamics and nonequilibrium fluctuation theorem at the same time. In fact, for a work distribution including initial coherence the first law of thermodynamics and nonequilibrium fluctuation theorem are mutually exclusive \cite{Llobet2017,Lostaglio2018,Xu2020}. This incompatibility sheds light on the crucial roles of quantum measurement and quantum coherence, and the necessity of explicitly considering an auxiliary system as a measurement apparatus rather than implicitly appealing to perform an energy measurement.

Already several works suggested the use of ancillas as the measurement apparatus for extracting the statistics of work. In Refs. \cite{Dorner2013,Mazzola2013} a Ramsey scheme using an auxiliary qubit was proposed to measure the work characteristic function. Its experimental realization with nuclear magnetic resonance was reported in Ref. \cite{Batalhao2014}. In Ref. \cite{Roncaglia2014} a different approach was taken in which a detector, for example the momentum of a quantum particle or a light mode \cite{Chiara2015}, is coupled to the system to extract directly the work probability distribution after a single measurement of detector position at final time. This scheme was recently realized with cold atoms \cite{Cerisola2017}. Although initial quantum coherence of the system gives rise to interference between the position eigenstates of the detector, its effect on work distribution can not be obtained by measuring detector position because position eigenstates are orthogonal to each other. For this, Solinas \textit{et al} focuses on the phase accumulated between the position eigenstates, and obtained quantum work distribution for arbitrary initial state by using full counting statistics method \cite{Solinas2015,Solinas2016,Solinas2017,Xu2018}. However, this resulting work distribution is a quasi-distribution because negative probability appears. The effect of initial quantum coherence on work probability distribution is still an open question. Since measuring the orthogonal states can not obtain the effect of initial quantum coherence, one may ask whether the initial quantum coherence can be recorded and measured by the nonorthogonal state. It is well known that coherent and squeezing states are typical nonorthogonal states \cite{Glauber1963,Cahill1969,Stoler1970,Stoler1971,Yuen1976}, and are widely used in quantum optics \cite{Scully1997}. Very recently, coherent and squeezing states are used to establish autonomous quantum Crooks theorem \cite{Kwon2019,Averg2018,Holmes2019,Mingo2019}. Coherent or squeezing state is the quantum extension of classical phase space \cite{Glauber1963,Cahill1969,Stoler1970,Stoler1971,Yuen1976}, the study of work distribution in coherent or squeezing state has a direct correspondence with the classical case. In present paper, we consider a harmonic oscillator with coherent or squeezing state as the detector to record work and study the role of quantum coherence in work probability distribution.

This paper is organized as follows: In the next section we give a brief review of the key concepts of coherent state and squeezing state. Based on coherent or squeezing state a unified framework of quantum work statistics for arbitrary initial state is given in Sec. III. As an application we consider a driven two-level system in Sec. IV. Finally, Sec. V closes the paper with some concluding remarks.

\section{Coherent state and squeezing state}

We begin by reviewing some key concepts of coherent state and squeezing state, allowing us to define the formalism that is used in the rest of our study. The details of coherent state and squeezing state can be found in the semina papers \cite{Glauber1963,Cahill1969,Stoler1970,Stoler1971,Yuen1976}. Coherent state is an eigenstate of annihilation operator $\hat{a}$ with the eigenvalue $\alpha$, i.e.,
\begin{equation}\label{coherent state_eigenvalue}
  \hat{a}|\alpha\rangle_a=\alpha|\alpha\rangle_a.
\end{equation}
Because $\hat{a}$ is not a Hermitian operator, $\alpha$ is a complex number. An expression of $|\alpha\rangle$ in terms of the number state $|n\rangle_a$ is given by
\begin{equation}\label{coherent state_definetion}
  |\alpha\rangle_a=e^{-|\alpha|^2/2}\sum_n\frac{\alpha^n}{\sqrt{n!}}|n\rangle_a.
\end{equation}
Since $|n\rangle_a=\frac{\hat{a}^\dag}{\sqrt{n!}}|0\rangle_a$ and $\exp(-\alpha^{\ast}\hat{a})|0\rangle_a=|0\rangle_a$, coherent state can then be expressed as
\begin{equation}\label{coherent state_dis}
  |\alpha\rangle_a=\hat{D}_a(\alpha)|0\rangle_a,
\end{equation}
where, $|0\rangle_a$ is the vacuum state with zero photon and
\begin{equation}\label{}
  \hat{D}_a(\alpha)=\exp(\alpha \hat{a}^{\dag}-\alpha^{\ast}\hat{a})
\end{equation}
is the displacement operator. The displacement operator $\hat{D}_a(\alpha)$ is a unitary operator, i.e., $\hat{D}_a^{\dag}(\alpha)=\hat{D}_a(-\alpha)=\hat{D}_a^{-1}(\alpha)$, and has the properties $\hat{D}_a^{\dag}(\alpha)\hat{a}\hat{D}_a(\alpha)=\hat{a}+\alpha$ and $\hat{D}_a^{\dag}(\alpha)\hat{a}^\dag\hat{D}_a(\alpha)=\hat{a}^\dag+\alpha^\ast$. Different coherent states are not orthogonal to each other, i.e.,
\begin{equation}\label{nonorthogonal}
  _a\langle\alpha|\alpha'\rangle_a=\exp\Bigl(-\frac{1}{2}|\alpha|^2+\alpha'\alpha^\ast-\frac{1}{2}|\alpha'|^2\Bigr),
\end{equation}
from which it follows that $|_a\langle\alpha|\alpha'\rangle_a|^2=\exp(-|\alpha-\alpha'|^2)$. If $\alpha$ and $\alpha'$ are quite different, i.e., $|\alpha-\alpha'|\gg1$, then $|\alpha'\rangle_a$ and $|\alpha\rangle_a$ are nearly orthogonal. Another consequence of their non-orthogonality is that the coherent states form an overcomplete basis, i.e.,
\begin{equation}\label{overcomplete}
  \frac{1}{\pi}\int d^2\alpha|\alpha\rangle_{a}{_a}\langle\alpha|=\mathbb{I}
\end{equation}
with $\mathbb{I}$ being the identity matrix and $d^2\alpha=d\mathrm{Re}(\alpha)d\mathrm{Im}(\alpha)$. One of the applications of the overcompleteness relation of coherent state is to calculate the trace, i.e.,
\begin{equation}\label{trace}
  \mathrm{Tr}[\hat{A}]=\frac{1}{\pi} \int{_a}\langle\alpha|\hat{A}|\alpha\rangle_a d^2\alpha,
\end{equation}
where $\hat{A}$ is an arbitrary operator.

Coherent state is closet to the classical state because it is a minimum-uncertainty state, in which the uncertainties of momentum $\hat{p}=\frac{1}{2i}(\hat{a}-\hat{a}^\dag)$ and coordinate $\hat{q}=\frac{1}{2}(\hat{a}+\hat{a}^\dag)$ satisfy $\Delta q\Delta p=1/4$ (in fact $\Delta q=\Delta p=1/2$). In this sense, coherent state can be used to establish the quantum extension of phase space. To be specific, the complex plane of $\alpha$ is corresponding to phase space, and $\mathrm{Re}(\alpha)$ and $\mathrm{Im}(\alpha)$ are corresponding to the classical coordinate and momentum respectively. We should noted that $\mathrm{Re}(\alpha)$ and $\mathrm{Im}(\alpha)$ are not the eigenvalues of $\hat{q}$ and $\hat{p}$, in fact $\mathrm{Re}(\alpha)=\langle \hat{q}\rangle$ and $\mathrm{Im}(\alpha)=\langle \hat{p}\rangle$ with $\langle \hat{q}\rangle$ and $\langle \hat{p}\rangle$ being the average of $\hat{q}$ and $\hat{p}$. In the quantum extension of phase space, the average of a microscopic observable $\hat{F}(\hat{a},\hat{a}^\dag)$ can be written as an integral of the product of a weight function $w(\alpha,\alpha^\ast)$ and a function $F(\alpha,\alpha^\ast)$ which refers to the operator $\hat{F}(\hat{a},\hat{a}^\dag)$,
\begin{equation}\label{F average}
 \langle \hat{F}(\hat{a},\hat{a}^\dag)\rangle\equiv\mathrm{Tr}[\hat{\rho} \hat{F}(\hat{a},\hat{a}^\dag)]=\int w(\alpha,\alpha^\ast)F(\alpha,\alpha^\ast)d^2\alpha,
\end{equation}
where $\hat{\rho}$ is the density matrix. This expression is similar to the phase space integrals in classical statistic mechanics. For the normal ordering operator $\hat{F}_N(\hat{a},\hat{a}^\dag)=\sum_{mn}c_{mn}(\hat{a}^\dag)^m\hat{a}^n$, its average can be expressed as
\begin{equation}\label{Normal}
  \langle \hat{F}_N(\hat{a},\hat{a}^\dag)\rangle=\int P(\alpha,\alpha^\ast)F_N(\alpha,\alpha^\ast)d^2\alpha,
\end{equation}
where
\begin{equation}\label{P}
  P(\alpha,\alpha^\ast)=\mathrm{Tr}[\hat{\rho}\delta(\alpha^\ast-\hat{a}^\dag)\delta(\alpha-\hat{a})]
\end{equation}
is the $P$ representation. In general, $P(\alpha,\alpha^\ast)$ is an extremely singular function. For the antinormal ordering operator $\hat{F}_A(\hat{a},\hat{a}^\dag)=\sum_{mn}c_{mn}\hat{a}^m(\hat{a}^\dag)^n$, its average can be expressed as
\begin{equation}\label{Antinormal}
  \langle \hat{F}_A(\hat{a},\hat{a}^\dag)\rangle=\int Q(\alpha,\alpha^\ast)F_A(\alpha,\alpha^\ast)d^2\alpha,
\end{equation}
where
\begin{equation}\label{Q}
  Q(\alpha,\alpha^\ast)=\frac{1}{\pi}{_a}\langle\alpha|\hat{\rho}|\alpha\rangle_a
\end{equation}
is the $Q$ representation.
For the symmetric ordering operator $\hat{F}_S(\hat{a},\hat{a}^\dag)=\sum_{mn}c_{mn}[\hat{a}^m(\hat{a}^\dag)^n+\hat{a}^n(\hat{a}^\dag)^m]$, its average can be expressed as
\begin{equation}\label{symmetric}
  \langle \hat{F}_S(\hat{a},\hat{a}^\dag)\rangle=\int W(\alpha,\alpha^\ast)F_S(\alpha,\alpha^\ast)d^2\alpha,
\end{equation}
where
\begin{equation}\label{W}
  W(\alpha,\alpha^\ast)=\frac{1}{\pi^2}\int d^2\nu\mathrm{Tr}\Bigl[\hat{\rho}\exp[\nu(\hat{a}^\dag-\alpha^\ast)-\nu^\ast(\hat{a}-\alpha)]\Bigr]
\end{equation}
is the Wigner-Weyl distribution. The Wigner-Weyl distribution is always a smooth function, but it can take negative values.

Squeezing state is defined as
\begin{equation}\label{squeezing state}
  |\alpha,\xi\rangle_a=\hat{S}_a(\xi)|\alpha\rangle_a
\end{equation}
with
\begin{equation}\label{squeezing operator}
  \hat{S}_a(\xi)=\exp\Bigl[\frac{1}{2}\Bigl(\xi^\ast \hat{a}^2-\xi (\hat{a}^\dag)^2\Bigr)\Bigr]
\end{equation}
being the squeezing operator, where $\xi=re^{i\theta}$ is an arbitrary complex number. Squeezing operator $\hat{S}_a(\xi)$ is a unitary operator, i.e., $\hat{S}_a^\dag(\xi)=\hat{S}_a^{-1}(\xi)=\hat{S}_a(-\xi)$. Squeezing state $|\alpha,\xi\rangle_a$ is also a minimum uncertainty state that $\Delta q\Delta p=1/4$, but the uncertainty of $\hat{q}$ or $\hat{p}$ can be $\Delta q<1/2$ or $\Delta p<1/2$ (depending on $\xi$), and this is the meaning of squeezing. From Eq. (\ref{coherent state_dis}), squeezing state can be rewritten as
\begin{equation}\label{squeezing-coherent}
  |\alpha,\xi\rangle_a=\hat{D}_b(\alpha)|0\rangle_b=|\alpha\rangle_b,
\end{equation}
where $\hat{D}_b(\alpha)=\exp(\alpha \hat{b}^{\dag}-\alpha^{\ast}\hat{b})$ is the displacement operator and $|0\rangle_b=\hat{S}_a(\xi)|0\rangle_a$ is the vacuum state in the new representation $\hat{b}=\hat{S}(\xi)\hat{a}\hat{S}^\dag(\xi)=\hat{a}\cosh r+\hat{a}^\dag e^{i\theta}\sinh r$ and $\hat{b}^\dag=\hat{S}(\xi)\hat{a}^\dag \hat{S}^\dag(\xi)=\hat{a}^\dag\cosh r+\hat{a} e^{-i\theta}\sinh r$. In this sense, the squeezing state in the representation $\hat{a}$ is the coherent state in the representation $\hat{b}$. The vacuum state in the representation $\hat{b}$ is also called squeezing vacuum state in the representation $\hat{a}$.

\section{Work measurement scheme using coherent or squeezing state}
Recently, a single measurement scheme that directly samples quantum work distribution was proposed in Ref. \cite{Roncaglia2014}, where the momentum of a quantum particle (auxiliary detector) is coupled to the system and then the position of the particle is shifted by an amount that depends on the energy change of the system. This scheme was experimentally realized by a cloud of ${^{87}}$Rb atoms \cite{Cerisola2017}, in which the system is represented by the Zeeman sublevels of ${^{87}}$Rb atom that behaves as a two-level system, the motional degree of freedom of the atom plays the role of the detector \cite{Machluf2013}. We note that the initial motional state is a wave-packet localised in position, it makes this single measurement scheme equivalent to TMP that the initial quantum coherence of the system is completely destroyed. In order to including the effects of initial quantum coherence, we consider that the auxiliary detector is supposed to be initially prepared in the squeezed vacuum state.

For the clarity of the discussion, we consider a harmonic oscillator as an auxiliary detector and review the main idea of the single measurement scheme in Ref. \cite {Roncaglia2014}. The Hamiltonian of the auxiliary detector is
\begin{equation}\label{}
  \hat{H}_a=\hbar\omega_a(\hat{a}^{\dag}\hat{a}+1/2),
\end{equation}
where $\omega_a$ is the oscillating frequency of the harmonic oscillator.
To determine the work, one needs the following five steps (see Fig. 1):
(1) at time $t<-\tau$, the auxiliary detector $a$ and the system $s$ are prepared in a product state $\hat{\rho}_a(0)\otimes\hat{\rho}_s(0)$.
(2) in order to know the initial system energy, at time $t=-\tau$, $a$ is coupled to $s$ with the Hamiltonian $\hat{H}_{sa}(\lambda_0)=-g\hat{p}\hat{H}_s(\lambda_0)$, where $g$ is the coupling strength. Notably, this interaction Hamiltonian does not influence the statistics of the initial system energy. The evolution of the total $s+a$ system is described by $\hat{U}_{sa}(\lambda_0)=\exp\{-i[\hat{H}_s(\lambda_0)+\hat{H}_a+\hat{H}_{sa}(\lambda_0)]\tau\}$. We assume that the time interval $\tau$ is short enough, i.e., the time interval $\tau$ is much shorter than the oscillating period of the harmonic oscillator and the characteristic time of the system (satisfying $\omega_a\tau\approx0$ and $\omega_s^{max}\tau\approx0$, where $\omega_s^{max}$ is the maximum transition frequency of the system). In this case, $\hat{U}_{sa}(\lambda_0)\approx\exp\{-i\hat{H}_{sa}(\lambda_0)\tau\}$.
(3) after the transient evolution of the total $s+a$ system, the coupling is removed at time $t=0$, and then a protocol is performed on $s$ with the work parameter being changed from its initial value $\lambda_{0}$ to the final value $\lambda_{t'}$.
(4) after that, $a$ is re-coupled with $s$ with the Hamiltonian $\hat{H}_{sa}(\lambda_{t'})=g\hat{p}\hat{H}_s(\lambda_{t'})$, and the transient evolution operator is $\hat{U}_{sa}(\lambda_{t'})\approx\exp\{-i\hat{H}_{sa}(\lambda_{t'})\tau\}$.
(5) perform the measurement (projective or weak) on the detector.
The information of the work is recorded on the measurement results of the detector, and one can obtain the work statistics through these measurement results.

\begin{figure}
\begin{center}
\includegraphics[width=8cm]{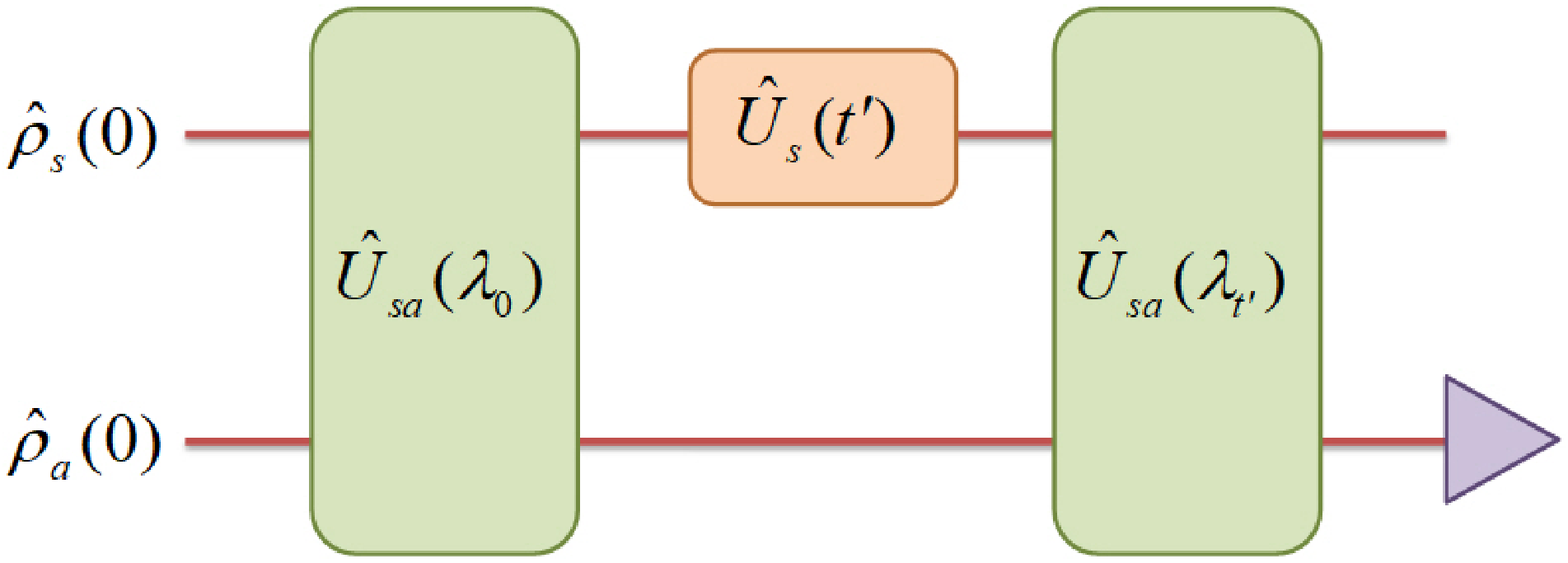}
\parbox{8cm}{\textbf{Fig. 1} \justifying (Color online) Schematic representation of work measurement scheme. Initially, the system and the detector is prepared in $\hat{\rho}_a(0)$ and $\hat{\rho}_s(0)$; Then, they are coupled with each other and evolved by $U_{sa}(\lambda_0)$ through a short time $\tau$; After the transient evolution, the coupling is removed and the work parameter being changed from its initial value $\lambda_{0}$ to the final value $\lambda_{t'}$ and the system is evolved by $U_s(t')$; The system and the detector is recoupled with each other and evolved by $U_{sa}(\lambda_{t'})$ through a short time $\tau$; Finally, a measurement (triangle) is performed on the detector.}
\end{center}
\end{figure}

In order to including the effects of quantum coherence, the auxiliary detector is supposed to be initially prepared in the squeezed vacuum state
\begin{equation}\label{}
  \hat{\rho}_a(0)=\hat{S}_a(r)|0\rangle_a{_a}\langle0|\hat{S}_a^{\dag}(r),
\end{equation}
where $|0\rangle_a$ is the vacuum state of auxiliary detector, $\hat{S}_a(r)=\exp\{\frac{r}{2}[\hat{a}^2-(\hat{a}^\dag)^2]\}$ is the squeezing operator, and $r\geq0$ is the squeezing strength (here the squeezing is performed only in real axis). Because $\hat{p}=\frac{1}{2i}(\hat{a}-\hat{a}^\dag)$, $\hat{U}_{sa}(\lambda_t)\approx\exp\{-\hat{H}_{s}(\lambda_t)\hat{a}^\dag+\hat{H}_{s}(\lambda_t)\hat{a}\}=\hat{D}_a(-\hat{H}_{s}(\lambda_t))$ (here we let $g\tau/2=1$), which can be understood as the displacement operator acting on the vacuum state of harmonic oscillator $|0\rangle_a$ can generate the coherent state depending on the energy level of the system, e.g. $\hat{D}_a(-\hat{H}_{s}(\lambda_t))|0\rangle_a\otimes|E^n_t\rangle=|-E^n_t\rangle_a\otimes|E^n_t\rangle$ where $|E^n_t\rangle$ is the $n$th eigenvectors of $\hat{H}_{s}(\lambda_t)$ with the corresponding eigenvalue $E^n_t$. $\hat{D}_a[\hat{H}_s(\lambda_{t})]$ acts on the squeezed vacuum state $\hat{S}_a(r)|0\rangle_a$ will generate the squeezed state whose parameter depends on the energy level of the system, e.g. $\hat{D}_a[-\hat{H}_{s}(\lambda_t)]\hat{S}_a(r)|0\rangle_a\otimes|E^n_t\rangle
=\hat{S}_a(r)D[-e^r\hat{H}_{s}(\lambda_t)]|0\rangle_a\otimes|E^n_t\rangle
=\hat{S}_a(r)|-e^rE^n_t\rangle_a\otimes|E^n_t\rangle=|-e^rE^n_t,r\rangle_a\otimes|E^n_t\rangle$.
From Eq. (\ref{squeezing-coherent}), we can see that the squeezing state $|-e^rE^n_t,r\rangle_a$ can be considered as the coherent state $|-e^rE^n_t\rangle_b$ in the representation $\hat{b}=\hat{a}\cosh r+\hat{a}^\dag\sinh r$.

After step (4), the state of auxiliary detector is
\begin{equation}\label{detector state}
  \hat{\rho}_{a}(t')=\sum_{lmn}\rho_{s}^{mn}(0) U_{lm}U^{\ast}_{ln}|\beta_r^{lm}\rangle_b{_b}\langle\beta_r^{ln}|,
\end{equation}
where, $\rho_{s}^{mn}(0)=\langle E^m_0|\hat{\rho}_s(0)|E^n_0\rangle$, $U_{lm}=\langle E_{t'}^l|\hat{U}_s(t')|E_0^m\rangle$ and $\beta_r^{lm}=e^r(E^l_{t'}-E^m_0)$. For closed system, the change of internal energy corresponds to work, i.e., $W=E^l_{t'}-E^m_0$. In other words, the work value is proportional to the real part of the parameter of coherent state $|\beta_r\rangle_b$, thus the work can be estimated by
\begin{equation}\label{}
  W=\mathrm{Re}(\beta_r)/e^r.
\end{equation}
According to $\overline{[\mathrm{Re}(\beta_r)]^n}=\sum_{m=0}^{m=n}C_n^m2^{-n}\pi^{-1}\int\langle\beta_r|(\hat{b}^\dag)^{n-m}\hat{\rho}_a(t')\hat{b}^m|\beta_r\rangle d^2\beta_r=2^{-n}\sum_{m=0}^{m=n}C_n^m\mathrm{Tr}\{\hat{b}^m(\hat{b}^\dag)^{n-m}\hat{\rho}_a(t')\}$, the $n$th moment of work is
\begin{equation}\label{}
\begin{split}
  \langle W^n\rangle&=\int \mathcal{P}(W)W^ndW \\
  &=\frac{e^{-nr}}{2^{n}}\sum_{m=0}^{m=n}C_n^m\mathrm{Tr}\{\hat{b}^m(\hat{b}^\dag)^{n-m}\hat{\rho}_a(t')\},
\end{split}
\end{equation}
where $C_n^m=\frac{n!}{m!(n-m)!}$. Because $\hat{b}^m(\hat{b}^\dag)^{n-m}$ is the antinormal ordering operator, $\mathrm{Tr}\{\hat{b}^m(\hat{b}^\dag)^{n-m}\hat{\rho}_a(t')\}=\int d^2\beta_r Q(\beta_r,\beta_r^\ast)\beta_r^m(\beta_r^\ast)^{n-m}$. In other words,
\begin{equation}\label{}
  \int P(W)W^ndW=\int\int d\mathrm{Im}(\beta_r)e^rQ(\beta_r,\beta_r^\ast)W^ndW,
\end{equation}
so the quantum work distribution is
\begin{equation}\label{}
  \mathcal{P}(W)=\int d\mathrm{Im}(\beta_r)e^rQ(\beta_r,\beta_r^\ast).
\end{equation}
According to Eq. (\ref{detector state}), the quantum work distribution is
\begin{equation}\label{work distribution}
\begin{split}
&\mathcal{P}(W)=\sum_{ln}P_nP_{l|n}\mathcal{N}\bigl(W\big\vert E^l_{t'}-E^n_0,\sigma\bigr) \\
      &+\sum_{\substack{lmn\\ m\neq n}}\varrho^{mn}_s(0)U_{ln}U^{\ast}_{lm}\mathcal{N}\Bigl(W\Big\vert E^l_{t'}-\frac{E^m_0+E^n_0}{2},\sigma\Bigr),
\end{split}
\end{equation}
where $P_n=\rho^{nn}_s(0)$ is the initial energy distribution, and $\mathcal{N}(W|\mu,\sigma)\equiv\exp\{-(W-\mu)^2/(2\sigma^2)\}/(\sqrt{2\pi}\sigma)$ is the normal distribution of $W$ with $\mu$ being the average value, $\sigma=\sqrt{2}\Delta q$ being the variance or the measurement error, and $\Delta q=\sqrt{\mathrm{Tr}[\hat{\rho}_a(t')\hat{q}^2]-\mathrm{Tr}[\hat{\rho}_a(t')\hat{q}]^2}=e^{-r}/2$ being the standard deviation of the position of detector. $\varrho^{mn}_s(0)=\rho^{mn}_s(0)\exp\{-(E^m_0-E^n_0)^2/(4\sigma^2)\}$ is the off diagonal element of the system density matrix after removing the coupling in step (3). The incoherent work value $W$ not only depends on the corresponding energy level transition with $E_m^{t'}-E_n^0=W$, but depends on all the transitions of the energy level by the Gaussian distribution form. In other words, the incoherent part of work probability distribution (\ref{work distribution}) is a coarse grained version of the result of TPM where Dirac delta functions have been replaced by Gaussians with measurement error $\sigma$. The transitions with the energy difference $E^m_{t'}-E^n_0=W$ gives the greatest contribution, and the more detuning between the work value $W$ and the energy level difference $E^m_{t'}-E^n_0$ the less contribution of this energy level transition.

From Eq. (\ref{work distribution}), it can be seen that quantum work not only depends on the initial energy distribution $P_m$ but also depends on the interference of the initial energy levels $\rho^{mn}_s(0)$ or quantum coherence. The quantum coherence interference of the initial energy levels $\rho^{mn}_s(0)$ or quantum coherence can be considered as the information that can be used to perform the work. Depending on energy difference ($E^m_0-E^n_0$) between the initial energy levels $|E^m_0\rangle$ and $|E^n_0\rangle$, their interference also contributes to the work by the Gaussian distribution form. The less the energy difference ($E^m_0-E^n_0$) between the energy levels, the more the contribution of their interference. For the same energy difference, the quantum coherence between the energy levels $|E^m_0\rangle$ and $|E^n_0\rangle$ with $E^l_{t'}-(E^m_0+E^n_0)/2=W$ gives the greatest contribution. The uncertainty of the position of the detector makes work distribution imprecise (can not recover the result of the TMP), but can survive some effects of initial quantum coherence from TPM. For the precise measurement, i.e., $\sigma=0$, the quantum work probability distribution is reduced to the result of TPM, i.e., $\lim_{r\rightarrow\infty}P(W)=\sum_{mn}P_nP_{m|n}\delta(W-(E^m_{t'}-E^n_0))$ in which only the transitions with the energy difference $E^m_{t'}-E^n_0=W$ contribute to the work with the value $W$.

After the Fourier transformation of the quantum work distribution $\chi=\int dWP(W)\exp(i\kappa W)$, the characteristic function of the quantum work distribution can be expressed as
\begin{equation}\label{chi}
\begin{split}
  \chi=e^{-\kappa^2\sigma^2/2}\mathrm{Tr}\Bigl[e^{i\kappa\hat{\mathcal{H}}_s(\lambda_{t'})}e^{-i\kappa \hat{\mathcal{H}}_s(\lambda_{0})/2}\hat{\varrho}_s(0)e^{-i\kappa\hat{\mathcal{H}}_s(\lambda_{0})/2}\Bigr],
\end{split}
\end{equation}
where $\hat{\mathcal{H}}_s(\lambda_{t})=\hat{U}_s^{\dag}(t)\hat{H}_s(\lambda_t)\hat{U}_s(t)$ is the system Hamiltonian at time $t$ in the Heisenberg picture. If the system is initially in the thermal equilibrium state $\hat{\rho}_{G}=\exp\{-\beta \hat{H}_s(\lambda_0)\}/Z(\lambda_0)$ with $\beta=1/(k_BT)$ being the inverse of the temperature, $Z(\lambda_0)=\mathrm{Tr}[\exp\{-\beta \hat{H}_s(\lambda_0)\}]$ being the partition function, and $k_B$ being the Boltzmann constant, we can obtain the modified Jarzynski equality (the quantum fluctuation relation) by letting $\kappa=i\beta$:
\begin{equation}\label{Jarzynski equality}
  \langle \exp\{-\beta W\}\rangle=\exp\{-\beta\Delta F\}\exp\biggl\{\frac{\beta^2\sigma^2}{2}\biggr\},
\end{equation}
where $\Delta F=k_BT\ln[Z(\lambda_{t'})/Z(\lambda_0)]$ is the variation of the Helmholtz free energy. This modified Jarzynski equality is consistent with the result of Ref. \cite{Watanabe2014}.

All the moments of the work done can be obtained by $\langle W^n\rangle=(-i)^n\partial^n\chi/\partial\kappa^n|_{\kappa=0}$. The average work is
\begin{equation}\label{W}
  \langle W\rangle=\mathrm{Tr}[\hat{H}_s(\lambda_{t'})\hat{\varrho}_s(t')]-\mathrm{Tr}[\hat{H}_s(\lambda_{0})\hat{\varrho}_s(0)],
\end{equation}
where $\hat{\varrho}_s(t')=\hat{U}_s(t')\hat{\varrho_s}(0)\hat{U}^{\dag}_s(t')$ is the evolution of the state after the first interaction with the detector. The interaction with the detector will inevitably change the system energy, and thus the average work we obtained includes the effects of the detector [$\langle W\rangle\neq\mathrm{Tr}[\hat{H}_s(\lambda_{t'})\hat{\rho}_s(t')]-\mathrm{Tr}[\hat{H}_s(\lambda_{0})\hat{\rho}_s(0)]$ with $\hat{\rho}_s(t')=\hat{U}_s(t')\hat{\rho}_s(0)\hat{U}^{\dag}_s(t')$]. If the system is initially in the classical state, it can not be influenced by the detector, i.e., $\hat{\varrho}_s(0)=\hat{\rho}_s(0)$, and the average work does not influenced by the measurement.

The second order moment of the quantum work is
\begin{equation}\label{W2}
  \langle W^2\rangle=\mathrm{Tr}\{\hat{\mathcal{H}}_s(\lambda_{t'})-\hat{\mathcal{H}}_s(\lambda_{0})]^2\hat{\varrho}_s(0)\}+\sigma^2.
\end{equation}
The work fluctuation can be expressed as
\begin{equation}\label{DELTAW}
  \delta W^2=\delta(\Delta\mathcal{H}_s)^2+\sigma^2
\end{equation}
with $\delta(\Delta\mathcal{H}_s)^2=\mathrm{Tr}\{\hat{\mathcal{H}}_s(\lambda_{t'})-\hat{\mathcal{H}}_s(\lambda_{0})]^2\hat{\varrho}_s(0)\}
-\mathrm{Tr}\{\hat{\mathcal{H}}_s(\lambda_{t'})-\hat{\mathcal{H}}_s(\lambda_{0})]\hat{\varrho}_s(0)\}^2$ being the variance of the change of the internal energy under the influence of the measurement. From Eq. (\ref{DELTAW}) it can be seen that the fluctuation of work for our measurement scheme is the variance of the change of the internal energy under the influence of the measurement plus the measurement error.

\begin{figure*}
\centering
\includegraphics[width=16cm]{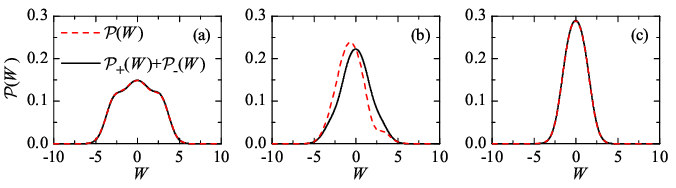}
\parbox{17.9cm}{\textbf{Fig. 2}. \justifying (Color online) Work distribution $\mathcal{P}(W)$ (red dashed curve) and the corresponding incoherent part $\mathcal{P}_+(W)+\mathcal{P}_-(W)$ (black solid curve) for (a) $t'=0.01\nu_0^{-1}$, (b) $t'=\nu_0^{-1}$ and (c) $t'=100\nu_0^{-1}$. For panels (a) and (c), $\mathcal{P}(W)$ and $\mathcal{P}_+(W)+\mathcal{P}_-(W)$ are coincide with each other. For all the panels, $\nu_{t'}=1.8\nu_0$, $\sigma=\nu_0$, $\beta=0.01\nu_0^{-1}$ and $\nu_0=1$.}
\end{figure*}

\section{Driven Two-Level System}
As an application, in this section we consider a nuclear spin system modulated by a radio frequency (rf) field in the transverse
($x$ and $y$) directions and investigate the effects of initial quantum coherence on work distribution. This nuclear spin system is widely used to experimentally investigate the quantum work distribution and fluctuation relation \cite{Batalhao2014}. The Hamiltonian of nuclear spin system is ($\hbar=1$)
\begin{equation}\label{Hs}
  \hat{H}_s(t)=\nu(t)\Bigl(\hat{\sigma}_x\cos\frac{\pi t}{2t'}+\hat{\sigma}_y\sin\frac{\pi t}{2t'}\Bigr),
\end{equation}
where $\hat{\sigma}_x=|1\rangle\langle0|+|0\rangle\langle1|$ and $\hat{\sigma}_y=-i|1\rangle\langle0|+i|0\rangle\langle1|$ are the Pauli operators, and $\nu(t)=\nu_0(1-t/t')+\nu_{t'}t/t'$ is the linear ramp of the rf field frequency over time $t'$, from $\nu_0$ to $\nu_{t'}$, $t\in[0,t']$. In order to investigate the effects of quantum coherence, we consider that the nuclear spin system is initially in the so-called coherent Gibbs state
\begin{equation}\label{}
  |\psi(0)\rangle=\sqrt{e^{-\beta\nu_0}/Z_0}|+\rangle+\sqrt{e^{\beta\nu_0}/Z_0}|-\rangle,
\end{equation}
where $|\pm\rangle=\tfrac{1}{\sqrt{2}}(|1\rangle\pm|0\rangle)$ is the eigenvector of $\hat{\sigma}_x$ with the corresponding eigenvalue $\pm1$, and $Z_0=e^{-\beta\nu_0}+e^{\beta\nu_0}$. It should be noted that $Z_0$ is the partition function of Gibbs state $\hat{\rho}_G(\nu_0)=e^{-\beta\hat{H}_s(0)}/Z_0$ with temperature $1/\beta$, and $|\psi(0)\rangle$ and $\hat{\rho}_G(\nu_0)$ are energetically indistinguishable because they have the same diagonal elements, in this sense, we call parameter $\beta$ in $|\psi(0)\rangle$ the ``temperature" or ``effective temperature".

According to Eq. (\ref{work distribution}), the work distribution performed by rf field on nuclear spin system is
\begin{equation}\label{}
  \mathcal{P}(W)=\mathcal{P}_+(W)+\mathcal{P}_-(W)+\mathcal{P}_c(W),
\end{equation}
where
\begin{equation}\label{}
\begin{split}
  \mathcal{P}_{\pm}(W)&=\frac{e^{\mp\beta\nu_0}}{Z_0}|U_{i,\pm}|^2\mathcal{N}\bigl(W\big\vert \nu_{t'}\mp\nu_0,\sigma\bigr) \\
  &+\frac{e^{\mp\beta\nu_0}}{Z_0}|U_{-i,\pm}|^2\mathcal{N}\bigl(W\big\vert -\nu_{t'}\mp\nu_0,\sigma\bigr)
\end{split}
\end{equation}
is the work distribution for the energy level $|\pm\rangle$, and
\begin{equation}\label{}
\mathcal{P}_c(W)=\tfrac{2e^{-\frac{\nu_0^2}{\sigma^2}}\mathrm{Re}(U_{i,+}U^\ast_{i,-})}{Z_0}
\Bigl[\mathcal{N}(W\vert\nu_{t'},\sigma)-\mathcal{N}(W\vert-\nu_{t'},\sigma)\Bigr]
\end{equation}
is the work distribution induced by the initial quantum coherence. In above equations, $U_{\pm i,\pm}=\langle\pm i|\hat{U}_s(t')|\pm\rangle$, $|\pm i\rangle=\tfrac{1}{\sqrt{2}}(|1\rangle\pm i|0\rangle)$ is the eigenvector of $\hat{\sigma}_y$ with the corresponding eigenvalue $\pm1$, and $\hat{U}_s(t')=\small\overleftarrow{T}\exp[-i\small\int_0^{t'}\hat{H}_s(t)dt]$ is the time evolution operator which needs to be calculated numerically. Fig. 2 shows work distribution for different evolution period $t'$. From Fig. 2 we can see that for the quench process $t'=0.01\nu_0^{-1}$ and adiabatic process $t'=100\nu_0^{-1}$, the initial quantum coherence has no effect on work distribution, but for the finite process $t'=\nu_0^{-1}$ the initial quantum coherence can make a significant contribution to work distribution. Besides, we also find that the work distribution performed by an adiabatic process is Gaussian (see Fig. 2(c)). At first sight, this result is not novel because it has been pointed out very recently \cite{Miller2019,Scandi2019}, but they are quite different. In Refs. \cite{Miller2019,Scandi2019}, the system is continuous and coupled to an environment, the Gaussian distribution is imported by the environment. But in our paper, the system is closed and finite, the Gaussian distribution is induced by work measurement. For the Gaussian work distribution, the work fluctuation is completely determined by the modified fluctuation-dissipation theorem:
\begin{equation}\label{MFD}
  \langle W_{irr}\rangle=\tfrac{1}{2}\beta(\delta W^2-\sigma^2)=\tfrac{1}{2}\beta \delta(\Delta\mathcal{H}_s)^2,
\end{equation}
where $\langle W_{irr}\rangle=\langle W\rangle-\Delta F$ is the irreversible work, $\Delta F=-\beta^{-1}\ln\tfrac{Z_{t'}}{Z_0}$ is the difference of free energy with $Z_{t'}=\exp[-\beta\nu_{t'}]+\exp[\beta\nu_{t'}]$. It should be noted that our fluctuation-dissipation theorem connects the irreversible work and the fluctuation of internal energy change after work measurement, which is different from the traditional fluctuation-dissipation theorem $\langle W_{irr}\rangle=\tfrac{1}{2}\beta\delta W^2$ in Ref. \cite{Jarzynski1997}. The traditional fluctuation-dissipation theorem is obtained by TPM in which initial quantum coherence is destroyed, our fluctuation-dissipation theorem is derived by the single measurement scheme, in which the initial quantum coherence is partially preserved by introducing the measurement error. And the measurement error is finally removed in our modified fluctuation dissipation theorem (\ref{MFD}).

Now we investigate the average work (the first moment of work) and the work fluctuation (the second moment of work). The average work can be expressed as
\begin{equation}\label{}
  \langle W\rangle=\langle W_+\rangle+\langle W_-\rangle+\langle W_c\rangle,
\end{equation}
where
\begin{equation}\label{}
  \langle W_{\pm}\rangle=\frac{e^{\mp\beta\nu_0}}{Z_0}\Bigl[|U_{i,\pm}|^2(\nu_{t'}\mp\nu_0)-|U_{-i,\pm}|^2(\nu_{t'}\pm\nu_0)\Bigr]
\end{equation}
is the average work for the energy level $|\pm\rangle$, and
\begin{equation}\label{}
\langle W_c\rangle=\frac{4}{Z_0}e^{-\nu_0^2/\sigma^2}\mathrm{Re}(U_{i,+}U^\ast_{i,-})\nu_T.
\end{equation}

\begin{figure}
\centering
\includegraphics[width=8cm]{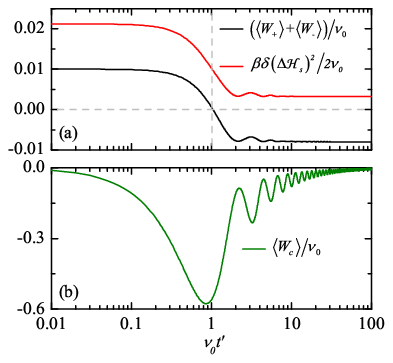}
\parbox{8cm}{\textbf{Fig. 3}. \justifying (Color online) (a) Incoherent work $\langle W_+\rangle+\langle W_-\rangle$ (black curve) and the fluctuation of the difference of internal energy $\tfrac{1}{2}\beta \delta(\Delta\mathcal{H}_s)^2$ (red curve) and (b) coherent work $\langle W_c\rangle$ (olive curve) as functions of evolution period $t'$. $\nu_{t'}=1.8\nu_0$, $\sigma=\nu_0$, $\beta=10^{-2}\nu_0^{-1}$ and $\nu_0=1$.}
\end{figure}

The incoherent part of average work $\langle W_+\rangle +\langle W_-\rangle$ is shown in Fig. 3(a). We can see that the amount of incoherent work is related to the driving velocity $1/t'$ or the driving period $t'$. If $\nu_0t'<1$, rf field will perform the positive work, but if $\nu_0t'>1$, rf field will perform negative work on the nuclear spin system. We also plot the fluctuation of the difference of internal energy $\tfrac{1}{2}\beta \delta(\Delta\mathcal{H}_s)^2$ (see red curves). The fluctuation of the difference of internal energy is
the biggest for the quench process, and it will be reduced by decreasing the driving velocity, until to the least for the adiabatic process. This can be understood as follows: Besides internal energy, quench process can also provide an extra energy that fluctuates the the system significantly; on the contrary, the adiabatic process only provides the internal energy, and have no extra energy to fluctuate the system. Interestingly, the behavior of $\langle W_+\rangle +\langle W_-\rangle$ is similar to that of fluctuation of the difference of internal energy $\tfrac{1}{2}\beta \delta(\Delta\mathcal{H}_s)^2$, which means that the fluctuation of the change of internal energy is mainly determined by incoherent work.

The coherent part of work is shown in Fig. 3(b). It can be seen that the coherent work for an adiabatic process $\nu_0 t'\sim\infty$ is zero. That is because adiabatic process can not induce energy level transition, thus initial quantum coherence can not be used to do work. The coherent work for the quench process $\nu_0 t'\sim0$ is zero too because the time of quench process is much shorter than the energy level transition time, initial quantum coherence has no time to contribute work. Only when the driving time matches the transition time of energy level, for the nuclear spin system we consider it is $t'\approx\nu_0^{-1}$, does initial quantum coherence contributes work significantly.

\section{Conclusions}
In this paper, we extended the traditional TPM by proposing a unified framework of quantum work statistics for arbitrary initial state (including quantum coherence). Specifically, we considered a harmonic oscillator with coherent state or squeezing state as a detector. The momentum of detector is coupled to the system Hamiltonian, and then the real part of the coherent state parameter of the detector will be linearly changed by the change of system energy, thus the work can be directly estimated by a single quantum measurement of the coherent state of the detector at final time. The resulting work probability distribution is positive and can be reduced to the result of TPM. To be specific, the incoherent part of our work probability distribution is the coarse grained version of the result of TPM where Dirac delta functions have been replaced by Gaussians. The uncertainty of our measurement scheme protects some effects of initial quantum coherence. Finally, we also considered a driven two-level system as an example, and found that only when the driving velocity matches the transition frequency of the system can initial quantum coherence play an important role.

\section{acknowledgement}
This work was supported by the National Natural Science Foundation of China (Grants No. 11705099, No. 11675017 and No. 11775019).

\end{document}